\documentstyle[12pt]{article}

\begin{document}
\setcounter{page}{0}

\newcommand{\rf}[1]{(\cite{#1})}
\newcommand{\R}{{\rm I\!R}}
\newcommand{\Z}{{\sf Z\!\!Z}}
\newcommand{\goto}{\rightarrow}
\newcommand{\Goto}{\Rightarrow}

\newcommand{\al}{\alpha}
\renewcommand{\b}{\beta}
\renewcommand{\c}{\chi}
\renewcommand{\d}{\delta}
\newcommand{\D}{\Delta}
\newcommand{\ve}{\varepsilon}
\newcommand{\f}{\phi}
\newcommand{\F}{\Phi}
\newcommand{\vf}{\varphi}
\newcommand{\g}{\gamma}
\newcommand{\ii}{\iota}
\newcommand{\G}{\Gamma}
\newcommand{\k}{\kappa}
\renewcommand{\l}{\lambda}
\renewcommand{\L}{\Lambda}
\newcommand{\m}{\mu}
\newcommand{\n}{\nu}
\newcommand{\r}{\rho}
\newcommand{\vr}{\varrho}
\renewcommand{\o}{\omega}
\renewcommand{\O}{\Omega}
\newcommand{\p}{\psi}
\renewcommand{\P}{\Psi}
\newcommand{\s}{\sigma}
\renewcommand{\S}{\Sigma}
\newcommand{\th}{\theta}
\newcommand{\vt}{\vartheta}
\renewcommand{\t}{\tau}
\newcommand{\vp}{\varpi}
\newcommand{\x}{\xi}
\newcommand{\z}{\zeta}
\newcommand{\ta}{\triangle}
\newcommand{\w}{\wedge}
\newcommand{\e}{\eta}
\newcommand{\Th}{\Theta}
\newcommand{\td}{\tilde}
\newcommand{\ep}{\epsilon}
\newcommand{\na}{\nabla}
\newcommand{\CW}{{\cal W}}
\newcommand{\CL}{{\cal L}}
\newcommand{\ha}{\frac{1}{2}}
\newcommand{\dt}{\frac{d}{dt}}
\newcommand{\be}{\begin{equation}}
\newcommand{\ee}[1]{\label{#1}\end{equation}}
\newcommand{\bE}{\begin{eqnarray}}
\newcommand{\eE}[1]{\label{#1}\end{eqnarray}}
\newcommand{\nn}{\nonumber}
\renewcommand{\thefootnote}{\fnsymbol{footnote}}

\newcommand{\siml}{\raisebox{-.6ex}{$\stackrel{<}{\displaystyle{\sim}}$}}
\newcommand{\simg}{\raisebox{-.6ex}{$\stackrel{>}{\displaystyle{\sim}}$}}
\newcommand{\ind}{\scriptscriptstyle}

\newcommand{\PR}{Phys. Rev. }
\newcommand{\PRL}{Phys. Rev. Lett. }
\newcommand{\NP}{Nucl. Phys. }

\newpage
\setcounter{page}{0}
\begin{titlepage}
\begin{flushright}
\hfill{KAIST-CHEP-94/08}\\
\vskip 2mm
\end{flushright}
\vspace{0.7cm}
\begin{center}
{\Large\bf Derivation of the Classical Lagrangian}
\vskip 3mm
{\Large\bf for the Relativistic Spinning Particle}

\vskip 1.5cm
{\bf Jin-Ho Cho \footnote{jhcho@chiak.kaist.ac.kr} and  Jae-Kwan Kim}
\vskip 0.3cm
{\sl Department of Physics\\
Korea Advanced Institute of Science and Technology\\
373-1 Yusung-ku, Taejon, 305-701, Korea}

\vskip 2.0cm

\end{center}

\setcounter{footnote}{0}

\begin{abstract}
The `classical' model for a massive spinning particle, which was recently
proposed, is derived from the isotropic rotator model. Through this
derivation, we note that the spin can be understood as the relativistic
extension of the isotropic rotator. Furthermore, the variables $t_\m$
corresponding to the $\p^*$ of the `pseudo-classical' model, are
necessary for the covariant formulation. The dynamical term for these
extra variables is naturally obtained and the meaning of the constraint
term $p^\s\L_{\s\n}+mt_\n =0$, which was recently shown to give
`quasi-supersymmetry', is clarified.
\end{abstract}

\end{titlepage}

\newpage
\renewcommand{\thefootnote}{\arabic{footnote}}
\baselineskip=18pt
Since the discovery of the spin effect by Stern-Gerlach experiment in 1922,
several attempts have been made to understand the `spin' concept.
Motivating the departure from the `classical' theory, `spin' has long been
understood as the `quantum' effect.

However, since the explanation of spin
via `Zitterbewegung' by Schr\"{o}dinger \cite{schro}, we still search for
some classical means to understand the spin \cite{han}. By the introduction
of supersymmetry \cite{rbe}, we came to know that spin is somewhat
`relativistic' effect rather than `quantum' concept. Those `pseudo-classical'
formulations use Grassmann variables to describe the spin degree of freedom.
We also can use c-number spinor for that \cite{barut}. Besides this
historical aspect, the relativistic spinning particle has been investigated
extensively due to its rich structure \cite{gat}.

In other direction, there have been some studies on the geometrical
construction of spin using group variables \cite{rba}. Furthermore, by the
modification of this `classical' model, some analogous features
(constraint structure and supersymmetry) with the pseudo-classical model
were found \cite{acho}. Moreover, Polyakov's spin factor \cite{pol} can be
obtained via BRST invariant path integral \cite{bcho}.

In this letter,
we clarify those concepts concerning the geometrical meaning of the
spin, the constraints and the variables $t^a$ corresponding to $\p^*$ of
the pseudo-classical model, through the derivation of the model recently
proposed in \cite{acho}. We start from the isotropic rotator model and
proceed to replace the variables, introduced to specify the orientation of
the rotator, with $SO(3)$ group variables. Further we give local $SO(3)$
symmetry to describe the variables with respect to the general rotating
frame. Finally, we transfer to the relativistic case.

To describe the non-relativistic free spinless particle, we use the usual
position variables $x^i$. The Lagrangian is then given by
\be
\CL_{posit}=\ha m \dot x ^2.
\ee{lapo}
For the non-relativistic isotropic rotator, in addition to the above
variables specifying the position of the center of mass, we introduce other
variables: the coordinates $r^i(t)$ for a point of the rotator body with
respect to the non-rotating frame\footnote{We should discriminate the
non-rotating frame, which has its origin at the center of mass and moves
with the rotator, from the observer-rest inertial frame where the observer
resides.}, to take into account the orientation of the rotator;
\bE
r(t)&=&O(t)R_0\nn\\
\dot r (t)&=&\dot O (t)O^{-1}(t)r(t)\nn\\
&\equiv&\vp (t)r(t)=w(t)\times r(t),
\eE{vel1}
where $R_0$ denote the coordinates for that point of the rotator body
with respect to the body-fixed frame, thus is time independent\footnote{In
the general rotating frame, $\dot r (t)=\dot O (t)R(t)+O(t)\dot R (t)
=O(t)(\dot R(t)+O^{-1}(t)\dot O (t)R(t))\equiv O(t)D_tR(t),$ where $D_t
=O^{-1}\dt O=\dt +O^{-1}\dot O (t)$ is the covariant derivative \cite{sat}.},
and $w(t)$ is the angular velocity of the rotator with respect to the
non-rotating frame. We can also define the corresponding quantities with
respect to the body-fixed frame;
\bE
\dot r (t)&=&O(t)O^{-1}(t)\dot O (t)R_0\nn\\
&\equiv&O(t)\O_0(t)R_0=O(t)(W_0(t)\times R_0)\nn\\
&\equiv&O(t)V_0(t).
\eE{vel2}
The time dependency of the coordinates $r(t)$ is entirely due to the
rotation $O(t) \in SO(3)$ of the body. Therefore, we can use $O(t)$ as
the orientation coordinates. In (\ref{vel1},\ref{vel2}) the following
isomorphism between $\{\R^3,\times \}$ and $\{so(3),[\,,\,]\}$ is to be
noted;
\be
\vp(t)=w^i(t)L_i,\,\,\O_0(t)=(W_0)^i(t)L_i,
\ee{iso}
where the basis $L_i$ for $so(3)$ is adjointly represented as
$(L_i)_{jk}=-\ep_{ijk}.$

The Lagrangian for this orientation part is given by
\bE
\CL_{orient}&=&\ha \int \r (r)\dot r ^2\,dr=-\ha \ii (t)w(t)^2\nn\\
&=&-\ha I_0 W_0^2,\nn\\
&=&-\ha <\O_0(t),\,I_0\O_0(t)>\nn\\
&=&-\ha <O^{-1}(t)\dot O (t),I_0O^{-1}(t)\dot O (t)>,
\eE{laor}
where $\r (r)$ is the mass density of the rotator and
$I_0=O^{-1}(t)\ii (t)O(t)$ is the moment of inertia tensor and $<\,\,,\,\,>
=-\ha Tr(\,\,,\,\,)$ is the canonical Killing form\footnote{$I_0$ is
symmetric with respect to this Killing form due to the trace property and
thus can be diagonalizable. For the isotropic rotator, it is proportional
to the identity.}. Finally the Lagrangian for the free isotropic
rotator becomes
\be
\CL_{rotator}=\ha m \dot x ^2-\ha <\tilde{S}_0(t),O^{-1}(t)\dot O (t)>,
\ee{laro}
where $\tilde{S}_0(t)\equiv I_0O^{-1}(t)\dot O (t)$ is the angular momentum
seen from the body-fixed frame. This system is very similar to the one in
\cite{rba}, where they construct the Lagrangian through topological
argument. Furthermore that dynamical term is nothing but the canonical action
that gives the symplectic structure on the group manifold \cite{alek}.

The system has the apparent global $SO(3)$ symmetry concerning the rotation
of the observer-rest inertial frame.
\be
x(t)\goto G\,x(t),\,\,r(t)\goto G\,r(t),\,\,O(t) \goto G\,O(t),
\ee{glo}
where $G \in SO(3)$ and we note that $R_0$ is invariant under this global
rotation. Together with this global symmetry we also expect a local $SO(3)$
symmetry. For $L(t)\in SO(3)$,
\bE
r(t)&=&O(t)L^{-1}(t)L(t)R_0\nn\\
    &=&O'(t)R(t)\,;\nn\\
O(t)&\goto&O(t)L^{-1}\nn\\
R_0&\goto&L(t)R_0.
\eE{loc}
This symmetry means the covariance of the system in the general
rotating frame (a frame with its origin at the center of mass and generally
rotating with respect to the body-fixed frame). We consider
this symmetry to be internal in the sense that $x(t)$
and $r(t)$ are invariant. However $\O_0(t)=O^{-1}(t)\dot O (t)$ does not
transform covariantly under this local internal symmetry;
\be
\O_0(t)\goto L(t)\O_0(t)L^{-1}(t)+L(t)\dot L ^{-1}(t).
\ee{affi}
Because of the second affine term, the Lagrangian (\ref{laro}) is not
invariant under the symmetry. To restore that local symmetry we replace
the body fixed frame with the general rotating frame.
\bE
R_0&\Goto &R(t)=g(t)R_0\nn\\
V_0(t)&\Goto &D_tR(t)=\dot R (t)+O^{-1}\dot O (t)R(t)\nn\\
&&=(\dot g (t)g^{-1}(t)+\O (t))R(t)\equiv(\CW (t)+\O (t))R(t).
\eE{gen}
Therefore $\O (t)$ is indeed the connection for this local symmetry and
$\CW (t)$ is the angular velocity of the rotator with respect to the
general rotating frame.

We are now ready to covariantize the Lagrangian for the orientation part;
\bE
\CL _{ORIENT}&=&\int \frac{\r (R)}{2}(D_tR)^2dR\nn\\
&=&\int \frac{\r (R)}{2}[(\CW +\O )R]^2dR\nn\\
&=&\int \frac{\r (R)}{2}<\CW +\O ,[\tilde{R},[\tilde{R},\CW +\O ]]>dR\nn\\
&=&-\ha <\CW +\O ,[\int -\r (R) (ad \tilde{R})^2\,dR](\CW +\O )>\nn\\
&=&-\ha <\CW +\O ,I(t)(\CW +\O )>,
\eE{cola}
where $\tilde{R}\equiv R^iL_i\in SO(3)$ and $I(t)=g(t)I_0g^{-1}(t)$ and
we omitted the explicit expression for the time dependence of the
variables. This orientation part also can be written in a compact form as
\be
\CL_{ORIENT}=-\ha <\tilde{S}(t),\CW (t)+\O (t)>
\ee{com}
with the definition $\tilde{S}(t)\equiv g(t)\tilde{S}_0(t)g^{-1}(t)
=S^i(t)L_i$.
For the local $SO(3)$ transformation the variables transform as;
\bE
r(t)&=&O(t)R(t)=O(t)L^{-1}(t)L(t)g(t)R_0;\nn\\
R'(t)&=&L(t)R(t),\nn\\
O'(t)&=&O(t)L^{-1}(t),\nn\\
g'(t)&=&L(t)g(t),\nn\\
\O '(t)&=&L(t)\O (t)L^{-1}+L(t)\dot L ^{-1}(t)\nn\\
\CW '(t)&=&L(t)\CW (t)L^{-1}+\dot L (t)L^{-1}(t).
\eE{gauge}
We see that $\CW (t)+\O (t)$ now transforms covariantly which assures that
local $SO(3)$ symmetry of the system. It should be noted that for the
covariant description we introduce other variables $g(t)\in SO(3)$.
Now the Lagrangian for the isotropic rotator reads as
\be
\CL _{ROTATOR}=\ha m\dot x ^2-\ha <\CW (t)+\O (t),I(t)(\CW (t)+\O (t))>.
\ee{lato}
We note that the Lagrangian (\ref{laro}) comes as a specific case;
\be
L(t)=g^{-1}(t)\Goto \CW '(t)+\O '(t)=(O(t)g(t))^{-1}\dt (O(t)g(t))
\equiv s^{-1}(t)\dot s (t).
\ee{bofix}
The system in \cite{rba} is obtained through the above gauge fixing
and with a choice of $S_0(t)$ in a specific direction\footnote{This is
possible because the variation of this $I_0s^{-1}(t)\dot s (t)$ part just
gives extra numerical factor that can be absorbed in the scale of that
specific direction.}. Further, the meaning of the $U(1)$ symmetry mentioned
there become clear with the view that it is the residual subsymmetry of this
local internal symmetry $SO(3)$.
For another specific case, we take the following `gauge fixing'.
\be
L(t)=O(t)\Goto \CW '(t)+\O '(t)=\dt (O(t)g(t))(O(t)g(t))^{-1}
\equiv \dot s (t)s^{-1}(t).
\ee{gfix}
This corresponds to the description of the system with respect to the
non-rotating frame.

Now we are in a position to consider the relativistic spinning particle.
For the relativistic description, we replace the variables with the
relativistic ones. Further, we do without those quantities
($I_0,R_0,I(t),R(t),\ii (t),r(t)$) which designate the extendedness of
the rotator since we now want to describe a point particle.
\bE
&t\goto \t ,O(t)\goto\L (\t),g(t)\goto \bar{\L}(\t),\nn\\
&\ha m \dot x ^2\goto p\dot x,\nn\\
&<\tilde{S}(t),\CW +\O >\goto
<\tilde{S}(t),\L^{-1}\dot{\L}+\dot{\bar{\L}}\bar{\L}^{-1}>,
\eE{chan}
where $\t$ is the invariant time parameter and changes the meaning of dot,
and $\L ,\bar{\L}\in SO(3,1)$. Now, with the basis $\s_{\m\n}$ of $so(3,1)$,
$\tilde{S}=S^{\m\n}\s_{\m\n}\in SO(3,1)$ and it concerns the
`internal' angular momentum. We use the first order formalism, accordingly
the following constraints should be taken into account.
\be
p^2+m^2=0,\,\,p=\L\bar{\L}m.
\ee{con}
Here, particle mass $m$ is defined in the `body-fixed frame'\footnote{Since
the point particle is bodiless this name is a little absurd, but we still
use that name avoiding perfusion of terminology.}. However it is
to be noted that the concept `non-rotating frame' is changed because four
dimensional rotation $\L\bar{\L}\in SO(3,1)$ includes boosting, which
makes the `non-rotating frame' depart from the `center of mass'. In this
relativistic case, we consider the `non-rotating frame' coincident with
the `observer-rest frame'. To define the spin that is intrinsic to the
particle, we fix the gauge so that $\tilde{S}$ is fixed in that
chosen rotating frame, like mass $m$ is the fixed momentum in the `body-fixed
frame'.

With the variables described with respect to this $\tilde{S}$ fixed
frame, the Lagrangian becomes
\be
\CL =p\cdot\dot x -\ha <\tilde{\l} ,\L^{-1}\dot \L
+\dot{\bar{\L}}\bar{\L}^{-1}>-N(p^2+m^2)-M\cdot\tilde{\l}(\L^{-1}p
-\bar{\L}m),
\ee{larel}
where the fixed $\tilde{S}$ is denoted as $\tilde{\l}$ and those constraints
(\ref{con}) are (global) Lorentz invariantly put into the Lagrangian with
the Lagrangian multipliers $N$ and $M_\m$. The $\tilde{\l}$ in the last
constraint term accounts for the vanishing of the term for the spinless
particle. Written in components, this Lagrangian leads to
\bE
\CL&=&p^\m\dot x _\m-\ha \l^{\m\n}\L^\s{}_\m\dot \L_{\s\n}
-\ha \l^{\m\n}\dot{\bar{\L}}_{\m\s}\bar{\L}_\n{}^\s\nn\\
&&-N(p^2+m^2)-M_\m\l^{\m\n}(p^\s\L_{\s\n}-\bar{\L}_{\n0} m).
\eE{lala}
It is to be noticed that only $\bar{\L}_{\n0}$ components of $\bar{\L}$
couple with other variables through the constraint terms; the components
$\bar{\L}_{\n i}$ can be completely decoupled away. This leaves the
following Lagrangian.
\bE
\CL=p^\m\dot x _\m-\ha \l^{\m\n}\L^\g{}_\m\dot \L_{\g\n}
-\ha \l^{\m\n}t_\m\dot t _\n
-N(p^2+m^2)-M_\m\l^{\m\n}(p^\s\L_{\s\n}+m t_\n),
\eE{lalu}
where $\bar{\L}_\n{}^0$ is denoted as $t_\n$. This is exactly the same
system as the classical model recently proposed in \cite{acho,bcho}.

We conclude this letter with some remarks. In three
dimension, we introduced two kinds of variables: $x^i (t)$ denote the
position whereas $r^i (t)$($O(t)$) specify the orientation. For the
covariant description under the local internal $SO(3)$, we introduce
other variables $g(t)$. Those variables are corresponding to the
observer-rest inertial frame, the non-rotating frame and the general rotating
frame
respectively. The latter two frames are internal in the sense that they are
pertaining to the rotator. However, relativistic extension shuffles the
observer-rest frame and the non-rotating frame because of the boosting.
We put the relation between these two frames through the second constraint
in (\ref{con}). This constraint does the role of bridge to connect those two
frame. Indeed, it corresponds to the constraint $p\cdot\p +m\p ^*=0$ of
the pseudo-classical formulation and generates `quasi-supersymmetry'
\cite{acho,bcho}.

The equations of motion can be easily derived by the
variational principle;
\bE
\d x^\m&\Goto&\frac{d}{d\t} p^\m=0\nn\\
\d \L&\Goto&\frac{d}{d\t}(x^\m p^\n -x^\n p^\m
+\L^\m{}_\r\L^\n{}_\s\l^{\r\s})=0.
\eE{moti}
With the spin component $\S^{\m\n}=\L^\m{}_\r\L^\n{}_\s\l^{\r\s}$, we can
see these are the right equations of motion for the relativistic spinning
particle. It can also be shown that the Dirac algebra for the momentum
$p^\m$ and the total angular momentum $J^{\m\n} =x^\m p^\n -x^\n p^\m
+\S^{\m\n}$ is isomorphic to the Poincar\'{e} algebra \cite{acho} and the
system gives the same spin factor as the one given by the pseudo-classical
model \cite{bcho}.

In this letter, we showed that the classical model for
the relativistic spinning particle can be obtained through the relativistic
extension of the isotropic rotator. We hope this derivation to be helpful
for the intuitive understanding of the `spin'.


\newpage

\begin{thebibliography}{99}

\bibitem{schro}  E. Schr\"{o}dinger and Sitzunber,
                 {\em Preuss. Akad. Wiss. Phys.-Math. Kl.}{\bf 24}
                 (1930) 418.
\bibitem{han} A. Hanson, T. Regge and C. Teitelboim,
              {\em Accademia Nazionale dei Lincei} (Roma, 1976).
\bibitem{rbe} F.A. Berezin and M.S. Marinov
             {\em Ann.Phys.}{\bf 104}(1977)336;
              L. Brink, P. Di Vecchia and P. Howe
              {\em Nucl. Phys. B}{\bf 118}(1977)76.
\bibitem{barut} A. O. Barut and Zanghi,
                {\em Phys. Rev. Lett.}{\bf 52}(1984)2009.
\bibitem{gat} A. Barducci, R. Caslbuoni, D. Dominici and R. Gatto,
              {\em Phys. Lett. B} {\bf 187} (1987)135;
              C. Batlle, J. Gomis and J. Roca,
              {\em Phys. Rev. D}{\bf 40}(1989)1950.
\bibitem{rba} A.P. Balachandran, G. Marmo, B.-S. Skagerstam and A. Stern,
             {\em Gauge Symmetries and Fiber Bundles}, (Springer-Verlag,
             Berlin, 1983).
\bibitem{acho} J.-H. Cho, S.J. Hyun and J.-K. Kim, ``A Covariant
              Formulation of Classical Spinning Particle'',
              hep-th/9302012 (To appear in {\em Mod. Phys. Lett. A}).
\bibitem{pol} A. M. Polyakov,
              {\em Les Houches, Session XLIX},(ed. E. Br\'ezin and J.
              Zinn-Justin,1988);
              J. Grundberg, T. H. Hansson, A. Karlhede and U. Lindstr\"om,
              {\em Phys. Lett. B}{\bf 218}(1989) 321;
              J. Grundberg, T. H. Hansson, and A. Karlhede,
              {\em Nucl. Phys. B}{\bf 347}(1990)420.
\bibitem{bcho} J.-H. Cho, S.J. Hyun and H.-j. Lee,``Polyakov's Spin Factor
for a Classical Spinning Particle via BRST Invariant Path Integral'',
              hep-th/9403103 (To appear in {\em Phys. Lett. B}).
\bibitem{sat} D. H. Sattinger and O. L. Weaver,
              {\em Lie Groups and Algebras with Applications to Physics,
              Geometry and Mechanic}, (Springer-Verlag, New York, 1986).
\bibitem{alek} A. Alekseev and S. Shatashivili,
               {\em Nucl. Phys. B}{\bf 323}(1989)719
\end{thebibliography}
\end{document}